\documentclass[a4paper,12pt]{article}
\usepackage{cite}
\usepackage{amsmath,amssymb,amsfonts}
\usepackage{algorithmic}
\usepackage{graphicx}
\usepackage{textcomp}
\usepackage{caption}

\newtheorem{theorem}{Theorem}

\newtheorem{proposition}{Proposition}

\newtheorem{definition}{Definition}
\newtheorem{corollary}{Corollary}
\newtheorem{assumption}{Assumption}

\def\BibTeX{{\rm B\kern-.05em{\sc i\kern-.025em b}\kern-.08em
    T\kern-.1667em\lower.7ex\hbox{E}\kern-.125emX}}

\begin{document}

\title{Dynamic generation and attribution of revenues in a video platform}

\author{Francisco Lopez-Navarrete$^*$,
\and Joaquin Sanchez-Soriano$^*$,
\and Oscar M. Bonastre\thanks{R.I. Center of Operations Research (CIO), Miguel Hernandez University of Elche, Avda de la Universidad s/n, Elche, 03202 Spain (e-mail: f.lopez.navarrete@gmail.com, joaquin@umh.es, oscar.martinez@umh.es) Corresponding author: Joaquin Sanchez-Soriano (e-mail: joaquin@umh.es).}}

\maketitle

\begin{abstract}
The consumption of online videos on the Internet grows every year, making it a market that increasingly generates a greater volume of income. This paper deals with a problem of great interest in this context: the allocation of the generated revenues in a video website between the website and the video creators. For this, we consider a dynamic model of the revenues generation. We will consider that revenue can come from two sources: through the pay-per-view system and through the insertion of advertisements in the videos. Then to study how to divide the revenues in a reasonable and fair way between the two parties, we consider a dynamic cooperative game that reflects the importance of each part in generating revenue. From this game, we determine how its Shapley value is and introduce other allocation rules derived from it. We provide a structure of algorithm to calculate the Shapley value and its derived rules. We show that the computational complexity of the algorithms is polynomial. Finally, we provide some illustrative examples and simulations to illustrate how the proposed allocation rules perform.
\end{abstract}


\section{Introduction}
\label{sec:introduction}
The great advance of communication technologies has allowed the creation of new paradigms of markets that until just a few decades ago were unimaginable. The growing and rapid spread of the use of Internet technologies along with wireless communications have opened the door to new business models based on the possibility of obtaining what is desired at any time and from anywhere. Some examples of success are Amazon, eBay, Alibaba, or YouTube that bill important amounts of money. Some of these business models are based on creating a ``marketplace'' that facilitates the interaction between two or more parties and that this interaction generates a profit. These markets can be studied as two-sided markets (see \cite{a1}, \cite{Rochet2006}, \cite{Roson2005}).

\begin{figure}
\begin{center}
\includegraphics[width=3in,height=2.5in,clip,keepaspectratio]{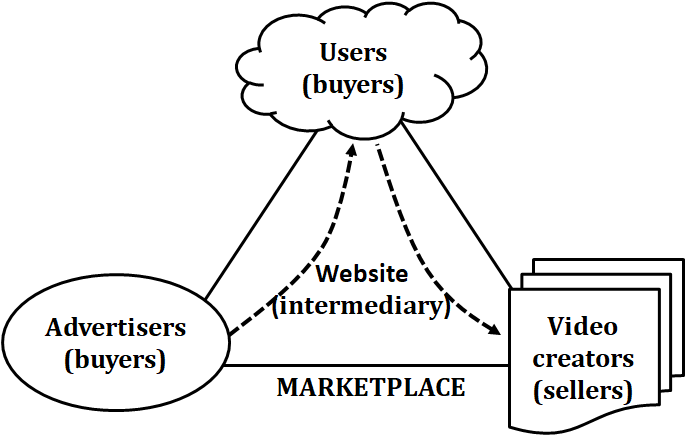}
\end{center}
\caption{Marketplace on a video on demand market.}
\label{marketplace}
\end{figure}

Princing, revenue allocation, resource allocation or Internet advertising are commonplace problems in the management of the new technologies. These problems are so relevant that we can not only find a large number of publications in the related literature but also numerous patents related to them. In particular, video markets are important to analyze because both the change in the way of consuming video content and the growing demand for videos on demand \cite{Gimpel2013}. Thus, we focus on video on demand two-sided markets. In these markets we basically have four types of agents. On the one hand, we have video creators who want to monetize their contents, users who want to watch videos, and advertisers who want users to watch their ads. Somehow, video creators are sellers, and users and advertisers are buyers. On the other hand, the marketplace where these agents are put in contact with each other so that transactions can occur is the video website, in which creators upload their contents, users watch videos, and advertisers insert their ads. In this sense, the video website is the intermediary between all parts. In Fig. \ref{marketplace}, we show a scheme of the basic structure of these two-sided markets.

For this type of market there are many different problems to analyze but we are only interested in how the revenue that this market can generate can fairly be distributed among the sellers (the creators of videos) and the intermediary (the web of videos). The other two vertices of the triangle would be the buyers: users and advertisers, but they do not take part in the distribution of the revenues. Why do we then not consider the buyers? On the one hand, users pay to the platform through subscriptions, pay-per-view or viewing the inserted ads. On the other hand, advertisers pay to the platform depending on the ads viewed and where they were inserted. Therefore, the platform receives all payments and has to compensate content creators. For this reason, we only consider the distribution problem between the platform and the creators. 

Problems of advertising, revenue allocation or revenue attribution in Internet businesses have aroused great interest both from an economic and research point of view. For this reason we can find numerous papers in the literature as well as patents. For example, how to maximize advertising revenues by delivering display ads on a website is studied in \cite{Roels2009}. A comprehensive survey on Internet advertising is provided in \cite{Yuan2012}. The design of efficient algorithms for deciding video-ad allocation is studied in \cite{Sumita2017}. A revenue-maximizing model that determines the optimal pay-per-view (PPV) price for each class, given the multi-class Video-on-Demand (VoD) services with differentiated qualities is analyzed in \cite{Kim2006}. Methods of revenue sharing, allocating monetization rights, revenue allocation related to different advertising and digital content provision business on Internet are proposed in a number of patents \cite{Heller2008,Fries2011,Redlich2012,Galai2019}. Therefore, the problem of allocating the revenue generated by video contents on Internet platforms or websites is a relevant topic to analyze.

Game theory has been applied in a multitude of fields, in particular to many engineering problems \cite{SS_2013}. Cooperative games have been applied to problems of resource allocation, revenue allocation or cost/benefit allocation. Some examples are the following. Pollution control \cite{Gimenez2016,Luqman2018,Gutierrez2018,Duro2020}, cloud technology problems \cite{Chen2017,Kim2018,Le2019}, transportation problems \cite{SS_2002,Guajardo2016,Algaba2019,Chen2020}, or wireless communication technology problems \cite{Niyato2006,Gozalvez2012,Lucas2012,Zhang2019,Tran2020} among many others. Therefore, the use of game theory is a good option to analyze the problem of revenue attribution in a video system on a web platform.

In \cite{Wang2012} the pricing and revenue sharing between one Internet access provider and IPTV providers who consider themselves independent monopolists are studied. For this, a three-stage non-cooperative price game model based on the theory of two-sided markets is used. In \cite{Kamiyama2010} the profit allocation problem among Internet service providers when they collaborate to provide VoD service is analyzed. In \cite{Lopez2019} a cooperative approach in which the revenue that one Internet TV service provider and many video content creators can obtain by means of cooperation is determined, and then  how to distribute that revenue among them is analyzed. In this paper, we study how to allocate revenue generated on a video website system. To do this, we consider the dynamic generation of revenue over time and analyze the influence on the generation of revenue of each movement of a user in their navigation on the website. In this sense, we approach the problem from the perspective of attribution methods. Since we consider a dynamic system and attribution problems, we will combine both dynamic cooperative game concepts \cite{Filar2000, Kranich2005} and attribution methods.

In our context of video contents in a web platform on Internet, an attribution method tries to detemine how credit for revenues is assigned to touchpoints in navigation paths in the system. Different attribution rules, including one based on the Shapley value \cite{Shapley1953},\footnote{For details about the relevance of the Shapley value in cooperative games and its applications in many different fields see, for instance, \cite{Roth1988,Algaba2019c}.} are considered by Google in the case in multi-channel funnels problems \cite{Googlea,Googleb}. However, we can find in the literature other attribution problems in different fields (see, for instance, \cite{Algaba2019b,SS_2020}). As mentioned before, in this paper we combine dynamic games and attribution methods to determine how to allocate the revenue generated on a video web platform between the different parties involved. In particular, to the best of our knowledge, the contributions of this paper to the literature are the following:

\begin{enumerate}
\item We model as dynamic games the generation of revenue in a video on demand (VoD) system on the Internet.
\item We determine the Shapley value of those games. This can be used to allocate the generated revenue between the different parties involved: platform and video contents creators. Therefore, we provide a way to determine how credit for revenues is assigned to touchpoints in navigation paths of users in the video web system.
\item Based on the Shapley value we introduce other revenue allocation rules. In particular, a new non-attenuated over rule and a family of attenuated over time rules. These rules also provide different ways  to determine how credit for revenues is assigned to touchpoints in navigation paths of users in the video web system.
\item We provide a polynomial algorithm for computing the allocation rules.
\end{enumerate}

In order to illustrate how the revenue allocation rules introduced in this paper perform, we carry out some computational experience by simulation. 

The rest of the paper is organized as follows. In Section II, we describe the video web system that we consider. In Section III, we present the mathematical model used to model the considered video web system. In Section IV, we introduce the dynamic games which we will use to determine how the revenue is generated when the different players collaborate. In Section V, we present the fairness properties that an allocation of the revenues should have in this context. In Section VI, we determine how the Shapley values of the introduced dynamic games   are and study their properties. Furthermore, other allocation rules based on the structure of the Shapley value are introduced and studied. In Section VII, an algorithm is presented to compute the Shapley value and the other allocation rules introduced in the previous section. Section VIII is devoted to illustrate how the Shapley value and the Shapley-like allocation rules perform. Section IX concludes. The proofs of the different statements and results in the paper can be found in the Appendix.


\section{System description}

In this paper we consider a video website in which content creators can upload their videos. When users enter the video website, they can generate revenue through two ways. On the one hand, users can pay-per-view the videos without having to watch advertisements and, on the other hand, users can watch the videos for free if they accept advertising inserted in them. In the first case, how the revenue is generated is evident: the user pays a certain price for watching the video, but in the second case we must bear in mind that advertisers will not pay the same if the ad is seen 5 seconds or 30 seconds. For example, on YouTube the revenue generated by advertising depends on several factors (https://www.youtube.com/intl/en-GB/yt/advertise/pricing/). Therefore, the same video may generate a different revenue from one time to another.

\begin{figure}
\begin{center}
\includegraphics[width=3.5in,height=2.5in,clip,keepaspectratio]{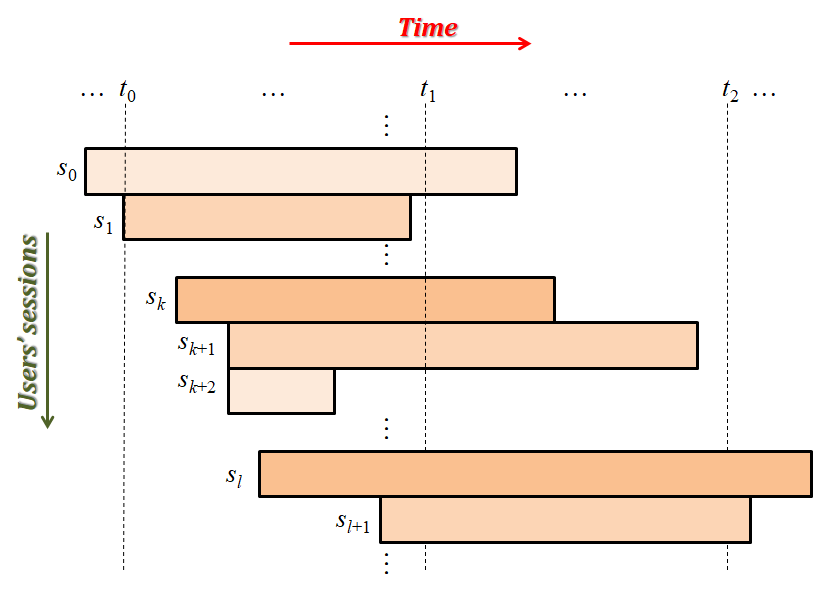}
\end{center}
\caption{Order of the user sessions and starting times.}
\label{sessions}
\end{figure}

Users will enter (or log in to) the video website in a certain order and instant of time, and will remain in the system for a while. So the dynamics of the system can be described in two dimensions, on the one hand, the order in which users enter (or log in) and, on the other hand, the exact time at which users start the session (see Fig. \ref{sessions}). Both dimensions are related in the sense that sessions $k$ and $k+1$ have started at moments of time such that $t_k \leq t_{k + 1}$. However, for practical purposes it is interesting to keep both dimensions for each session initiated by the users of the video website.

\begin{figure}
\begin{center}
\includegraphics[width=3.5in,height=2.5in,clip,keepaspectratio]{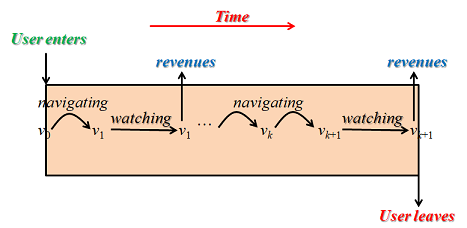}
\end{center}
\caption{Navigation of a user during a session.}
\label{navigation}
\end{figure}

When a user enters the video website, she will navigate from one video to another, she will view or not some videos and will leave the system after a while. During her whole session revenues will be generated by means of pay-per-view and/or advertising (see Fig. \ref{navigation}).

If we analyze the process of revenue generation we can consider that there are three types of elements involved. On the one hand, the website where the videos are uploaded, on the other hand, the videos that are viewed and, finally, the element that attracts the user, which can be a specific video or the website itself. These three elements are essential so that revenues can be generated, since if one of them is missing, it is not possible. If there is no website, there are no videos, if there are no videos, there is neither pay-per-view nor advertisers that pay to insert their ads, and if there is nothing that attracts users to the system, there are no users, and without users there is neither pay-per-view nor advertisers interested in inserting their ads in videos.

In addition to the above, the video website can offer a recommendation service and a search service, uploaded videos are usually grouped by channels and users can subscribe to those channels. All of the latter can also be interesting to take into account when considering how the revenues of the system are generated.


\section{Mathematical model}

In this paper, we will consider a mathematical model to describe the generation of revenue in a video on demand (VoD) system on the Internet. This will be a discrete model in which does not take exactly into account how revenue is generated over time. Thus, the model is discrete event dynamic.

We shall denote by $W$ the video website, by $U$ the universe of all possible users, and by $C$ the set of all possible channels in the video website $W$ in any time. Note that both the universe of users and the universe of channels are open, in the sense that they are neither limited to a set of specific users nor a particular video channels nor each video channel to have a fixed number of videos. Therefore, this reflects in some way that the users, the channels, and the videos on the website can change dynamically along the time.

We now denote by $S(t)$ the set of all sessions started and ended by users in $U$ from the beginning until time instant $t$, by $S(t_1, t_2)$ the set of all sessions ended by the users in $U$ between time instants $t_1$ and $t_2$, i.e. in the time interval $]t_1,t_2]$, such that $t_1 < t_2$. Note that $S(t)=S(0,t)$ and $S(t_1, t_2)=S(t_2)-S(t_1)$. Given a session $s$, we consider the following information:
\begin{itemize}
\item $a(s)$ is the element in $W \cup C$ that attracted the user who started session $s$.

\item $v(s)$ is the set of videos in $W$ that the user viewed in session $s$, of course, each video will belong to a channel in $C$, but different videos may belong to the same channel.

\item $r(i,s)$ is the revenue obtained from video $i \in v(s)$.

\item We can consider that the video website consists of three agents: $w_p$ is the platform itself, $w_r$ is the recommendation service, and $w_s$ is the search service. The two last agents could also generate revenues during a session because some ads could be inserted or poped up during the recommendation or searching process, if not their revenue associated would be simply zero. Thus $r(w_r,s)$ is the revenue obtained when the user is looking at suggestions of videos, and $r(w_s,s)$ is the revenue obtained when the user searches.

\item $R(s)$ is the whole revenue obtained in session $s$ and is given by

\begin{equation}
R(s)=r(w_r, s)+r(w_s, s)+\sum_{i \in v(s)} r(i, s).
\end{equation}

\end{itemize}
 
According to the notation above, the total revenue obtained by the video website during the period time window $]t_1,t_2]$ is given by

\begin{equation}\label{revenue}
R(t_1,t_2)=\sum_{s \in S(t_1,t_2)}R(s).
\end{equation}

Therefore, we have described the revenue generation in terms of the different sessions initiated by the users, and for each session we have considered the revenue generated for each viewed video, the recommendation service, and the search service. Likewise, Eq. \ref{revenue} means that we consider that a session is only monetized when finished.


\section{Dynamic games}
One way to allocate the revenue between the stakeholders is to use cooperative game theory. For this, first we have to identify who are the players, and second, we have to define the characteristic function. In our case, it seems reasonable that the players will be the video website and the video channels, i.e., $N=W \cup C$. Furthermore, we will also divide the video website into three agents: the platform itself, the search service and the recommender service, i.e., $W=\{w_p,w_s,w_r\}$. The next step is to define the characteristic function which measures, in this case, the revenue generated by a coalition of players. In order to do this, we have to consider some reasonable assumptions. The first assumption is the following:
\medskip
\begin{assumption}\label{a1}
If the platform does not belong to a coalition, then the revenue generated by that coalition is zero.
\end{assumption}
\medskip
Assumption \ref{a1} seems very reasonable, because if there is no platform, then no revenues can be generated on the video website. Moreover, when a user begins a session in the system, this session is going to consist of a chain of events over time as depicted in Fig. \ref{navigation}. How we consider the relevance of each event of the chain will be very important to determine the characteristic function. We must make some assumption about the relevance of the events of a user's navigation chain (i.e., a session). Depending on this assumption about the relevance of the events of a session we will obtain a game or another. In principle, we have two alternatives.

\medskip
\begin{assumption}\label{a2-all_or_nothing}
If an event is removed from a session, then after that event no revenues are generated.
\end{assumption}
\medskip
\begin{assumption}\label{a3-nothinghappens}
If an event is removed from a session, then only the revenues associated with that event are removed.
\end{assumption}
\medskip

Assumption \ref{a2-all_or_nothing} implies that if an event is removed then the session ends immediately and no additional revenues can be generated after that moment. However, Assumption \ref{a3-nothinghappens} means that if an event is removed from a session, the session continues but simply omitting that event in the revenue generation. Therefore, both assumptions assume somehow the two extremes. On the one hand, Assumption \ref{a2-all_or_nothing} is the case in which every event of a session is given maximum importance in revenue generation. And, on the other hand, Assumption \ref{a3-nothinghappens} is the case where no relevance in extra revenue generation is given to each event in a session.

Once we have fixed the players, and some reasonable assumptions on revenue generation, we can define the characteristic function taking into account the mathematical model of revenue generation to be used: discrete or continuous.

In order to simplify the notation, hereafter we will consider that a session $s$ consists of a finite sequence of $n_s +1$ events, 
\begin{equation}\label{eventssession}
\{e_0(s), e_1(s),...,e_{n_s}(s)\},
\end{equation}
and $P$ is a map from the set of all possible events to $N = W \cup C$, i.e., $P$ maps each event to its owning player. In order to simplify the notation, we write $P(e_k(s))=P_k(s)$. Furthermore, we assume that $P(e_0(s)) =P_0(s) =w_p$ and $P(e_k(s))=P_k(s) \in \{w_s,w_r\} \cup C$, for all $k \in \{1,...,n_s\}$. This means that we consider a session always starts on the platform, and then different videos belonging to possibly distinct channels are visited, as well as the search service and the recommendation service, until the session ends after a finite number of events. Moreover, the identification of videos with the video channel to which it belongs seems reasonable to define the corresponding game since the players are the video channels. In addition, we denote by
\begin{equation}\label{firstk+1events}
E^k(s)=\{e_0(s), e_1(s),...,e_k(s)\}
\end{equation}
and
\begin{equation}\label{firstandk+1event}
\widetilde{E}^k(s)=\{e_0(s), e_k(s)\}
\end{equation}
the set of the first $k+1$ events in session $s$ and the set of the first and $k+1$-th events in session $s$, respectively. Finally, we denote by
\begin{equation}\label{numplayersEk}
P(E^k(s)) =\{P_0(s),P_1(s),...,P_k(s)\} \subset N,
\end{equation}
and $n^p_k(s)=|P(E^k(s))|$ is the number of listing players within the events set $E^k(s)$; and we denote by
\begin{equation}\label{numplayersEk}
P(\widetilde{E}^k(s)) =\{P_0(s),P_k(s)\}\subset N,
\end{equation}
and $\widetilde{n}^p_k(s) =|P(\widetilde{E}^k(s))|$. Note that $\widetilde{n}^p_k(s)$ is always $2$ except when $k=0$ that it is $1$.

\subsection{Some preliminary definitions}\label{preliminar}
Before starting with the game theoretical models we shall use to allocate the revenues, we introduce some game theory definitions and concepts we will use throughout this paper.

\medskip

\begin{definition}
A dynamic cooperative game is a pair $(N,V)$, where $N$ is a finite set of players and $V$ is a function, called the characteristic function of the game, from $2^{N} \times T\times T$ to $\mathbb{R}$, such that $2^{N}$ is the set of all possible coalitions in $N$ and $T=[0,+\infty[$ is the timeline. Furthermore, $V$ satisfies the following properties:
\begin{enumerate}
\item $V(\varnothing,t_1,t_2) =0, \forall t_1 \leq t_2$.
\item $V(F,t_1,t_2)=0, \forall F \subset N \text{ and } \forall t_1 \geq t_2 \geq 0$.
\item $V(F,t,t)=0, \forall F \subset N \text{ and } \forall t \in \mathbb{R}_+$.
\end{enumerate}
\end{definition}

The first property means that the empty set does not generates any revenue. The second property means that the moving backwards does not generate any revenue. Finally, the third property means that revenues cannot be generated with no time. This three properties are absolutely reasonable in realistic settings.

The characteristic function $V$ evolves along the time, that means $V(F,t_1,t_2)$ represents what the coalition $F \subset N$ can obtain in the time window $]t_1,t_2]$. Therefore we consider intervals of time which are open on the left and closed on the right. This is reasonable because it makes no sense to consider a session that just ends when the time window begins, taking into account that the value of just one time point is zero.

When no confusion is possible, we denote $V(F,0,t)$ by simply $V(F,t)$. The characteristic function $V$ represents the revenue that the players in a coalition can obtain by themselves in a time window.

An interesting property for dynamic cooperative games is convexity, because it implies that cooperation is certainly profitable for all agents.

\medskip

\begin{definition}
A dynamic cooperative game $(N,V)$ is timewise convex, if the cooperative game $(N, V(\cdot,t))$ is convex for every $t \in T$, i.e., $\forall i \in N \text{ and } \forall F \subset G \subset N \backslash \{i\}$, the following holds
\begin{equation}
V(F \cup \{i\},t) - V(F,t) \leq V(G \cup \{i\},t) - V(G,t).
\end{equation}
\end{definition}

Another interesting property related to the time dimension of the game is time separability. This property says that the value of a dynamic game can be obtained by decomposing time windows into smaller intervals of time.
\medskip
\begin{definition}
A dynamic cooperative game $(N,V)$ is time separable, if $\forall t_1,t_3 \in T \text{ and } \forall F \subset G \subset N$, the following holds
\begin{equation}
V(F,t_1,t_3) = V(F,t_1,t_2) + V(F,t_2,t_3), \forall t_2 \in ]t_1,t_3].
\end{equation}
\end{definition}

Given a dynamic cooperative game, the main objective is to determine how to allocate the revenue generated among the stakeholders. Therefore, we need to establish what we mean by an assignment in this dynamic environment.
\medskip
\begin{definition}
Given a game $(N,V)$ and a time window $]t_1,t_2]$, an allocation is a vector $x^{]t_1,t_2]} \in \mathbb{R}^{N}$, such that $x^{]t_1,t_2]}(N) = \sum_{i\in N}x_i^{]t_1,t_2]}=V(N,t_1,t_2)$. We denote the set of all allocations of a time window $]t_1,t_2]$ by $A^{\left(N,V(\cdot,t_1,t_2)\right)}$ 
\end{definition}

\medskip

We will simply write $x^{t}$ for an allocation associated with the time window $]0,t]$.

\medskip

\begin{definition}\label{timevalues}
A time window value is a function $\Phi$ that assigns to each game $(N,V)$ a collection of allocations $\Phi(N,V)=\left\{x^{]t_1,t_2]}\right\}_{]t_1,t_2] \subset ]0,+\infty[}$, one allocation for each time window. And a timewise value is a function $\widehat{\Phi}$ that assigns to each game $(N,V)$ a collection of allocations $\widehat{\Phi}(N,V)=\left\{x^t \right\}_{t \in \mathbb{R}_+}$, one allocation for each time moment.
\end{definition}

A time window (resp. timewise) value is actually a revenue allocation rule that determines how much of the revenue has to be given to each player in each time window (resp. until a certain time). Therefore, from now on we will use indistinctly value or allocation rule to refer to this concept. The most relevant value in game theory is the Shapley value \cite{Shapley1953,Roth1988,Algaba2019c,Filar2000}. For dynamic cooperative games we can define the time window and the timewise Shapley value as follows:

\medskip

\begin{definition}\label{twsh}
The time window Shapley value, $Sh$, assigns to each game $(N,V)$ a collection of allocations $Sh(N,V)=\left\{Sh^{]t_1,t_2]}(N,V)\right\}_{]t_1,t_2] \subset ]0,+\infty[}$, i.e., the Shapley value for each time window game. In formula, for each $i \in N$, it is given by
\begin{align}
&Sh_i^{]t_1,t_2]}(N,V)= \nonumber \\
&\sum_{F \subset N\backslash \{i\}}{\textstyle\frac{|F|!|N\backslash (F\cup \{i\})|!}{|N|!}}\left(V(F\cup \{i\},t_1,t_2)-V(F,t_1,t_2)\right)
\end{align}
And the timewise Shapley value, $\widehat{Sh}$ assigns to each game $(N,V)$ a collection of allocations $\widehat{Sh}(N,V)=\left\{ Sh^t(N,V) \right\}_{t \in \mathbb{R}_+}$, i.e., the Shapley value for each time moment game. In formula, for each $i \in N$, it is given by
\begin{align}
&Sh_i^t(N,V)= \nonumber \\
&\sum_{F \subset N\backslash \{i\}}{\textstyle\frac{|F|!|N\backslash (F\cup \{i\})|!}{|N|!}}\left(V(F\cup \{i\},t)-V(F,t)\right)
\end{align}
\end{definition}

If in Definition \ref{timevalues} instead of considering only one allocation for time window (resp. time moment), we consider a set of allocations, then we obtain a time window solution (resp. timewise solution). One of the most relevant solutions in cooperative game theory is the core \cite{Gillies1953,Filar2000,Kranich2005}, because it provides the set of all coalitional stable allocations of the game. The core for a time window and for a time moment are given by
\begin{align}
&C(N,V(\cdot,t_1,t_2))=\{x^{]t_1,t_2]}\in A^{\left(N,V(\cdot,t_1,t_2)\right)}|x^{]t_1,t_2]}(F) \geq \nonumber \\
& V(F,t_1,t_2), \forall F \varsubsetneq N \text{ and } x^{]t_1,t_2]}(N) \geq V(N,t_1,t_2)\}
\end{align}
\begin{align}
&C(N,V(\cdot,t))=\{x^{t}\in A^{\left(N,V(\cdot,t)\right)}|x^{t}(F) \geq V(F,t), \nonumber \\
&\forall F \varsubsetneq N  \text{ and } x^{t}(N) \geq V(N,t) \}
\end{align}

Finally, given two sets $K, F \subset N$, we define a function $\chi$ like the Kronecker delta but for the inclusion relationship between two sets  as follows:

\begin{equation}
\chi(K,F)=
\begin{cases}
1, & if K \subset F, \\
0, & otherwise. 
\end{cases}
\end{equation}


\subsection{Discrete event dynamic games}\label{discreteeventgames}
First, the set of players is $N=W \cup C$, and under Assumption \ref{a1} and Assumption \ref{a2-all_or_nothing}, the characteristic function, $V$, will be given for every coalition $F \subset N$ and every time window $]t_1,t_2]$ by

\begin{equation}\label{DD12a}
V(F,t_1,t_2)=\sum_{s\in S(t_1,t_2)} R(s|F)
\end{equation}
where
\begin{equation}\label{DD12b}
R(s|F) =\sum_{k\leq max\{h\in \mathbb{Z}|P(E^h(s)) \subset F\}}r(e_k(s)).
\end{equation}

The family of all of these dynamic games with set of player $N$ is denoted by $DD_{12}^{N}$, and the family of all these games is simply denoted by $DD_{12}$.

However, under Assumption \ref{a1} and Assumption \ref{a3-nothinghappens},  the characteristic function, $\widetilde{V}$, will be given for every coalition $F \subset N$ and every time window $]t_1,t_2]$ by

\begin{equation}\label{DD13a}
\widetilde{V}(F,t_1,t_2)=\sum_{s\in S(t_1,t_2)} \widetilde{R}(s|F)
\end{equation}
where
\begin{equation}\label{DD13b}
\widetilde{R}(s|F) =\sum_{k : P(\widetilde{E}^k(s)) \subset F}r(e_k(s)).
\end{equation}

The family of all of these dynamic games with set of player $N$ is denoted by $DD_{13}^{N}$, and the family of all these games is simply denoted by $DD_{13}$.

Next we study properties of these games related to both dimensions, coalitional and time-wise. First, we show that these games are convex.

\medskip
\begin{proposition}\label{convexgames}
The games in $DD_{12}$, and $DD_{13}$ are timewise convex.
\end{proposition}

This result is important because it means that cooperation is profitable for all agents involved. Furthermore, the well-known Shapley value belongs to the core of the game \cite{Shapley1971}, what implies that the Shapley value is coalitionally stable and then a good candidate to be used in the allocation of revenues.

The next proposition shows the properties of these games related to the time component of their definition.
\medskip
\begin{proposition}
The games in $DD_{12}$, and $DD_{13}$ are time separable.
\end{proposition}


\section{Fairness properties}\label{properties}
In order to study what kind of allocation rules are more suitable, we should give a collection of properties which are meaningful for the problem at hand. Meaningful can refer to fairness, technical maneuverability, stability or other criteria. In our case, we are interested in the three mentioned criteria: fairness, technical simplicity and stability. Fairness is required because it is desirable that all the agents involved be treated fairly. Technical maneuverability so that the rule is simple to calculate, i.e., that it is effectively implementable. Stability as a criterion of consistency, in the sense that no complaints are generated and therefore the system can work in the long term.

The properties are given for time window values but they can {\it mutatis mutandis} be adapted to timewise values.

\begin{itemize}
\item Efficiency (EFF). A time window value $\Phi$ is efficient if for all game $(N,V)$, and for all time window $]t_1,t_2]$, 
\begin{equation}
\sum_{i\in N}x_i^{]t_1,t_2]} = V(N,t_1,t_2).
\end{equation}

\item Symmetry (SYM). A time window value $\Phi$ is symmetric if given a dynamic cooperative game $(N,V)$, such that there are $i,j\in N$ such that for all time window $]t_1,t_2]$, 
\begin{equation}
V(F\cup\{i\},t_1,t_2)=V(F\cup\{j\},t_1,t_2), \forall F\subset N\backslash \{i,j\},
\end{equation}
then it holds that
\begin{equation}
x_i^{]t_1,t_2]}=x_j^{]t_1,t_2]}, \forall ]t_1,t_2] \subset ]0, +\infty[.
\end{equation}

\item Null player (NP). A time window value $\Phi$ satisfies the null player property, if given a dynamic cooperative game $(N,V)$, such that there is $i\in N$ such that for all time window $]t_1,t_2]$,
\begin{equation}
V(F\cup\{i\},t_1,t_2)=V(F,t_1,t_2), \forall F\subset N\backslash \{i\},
\end{equation}
then it holds that
\begin{equation}
x_i^{]t_1,t_2]}=0, \forall ]t_1,t_2] \subset ]0, +\infty[.
\end{equation}

\item Monotonicity (MON). A time window value $\Phi$ is monotonic if given games $(N,V)$ and $(N,V')$ such that $\forall ]t_1,t_2] \subset ]0,+\infty[$,
\begin{equation}
V(F,t_1,t_2)= V'(F,t_1,t_2), \forall F \subsetneq N,
\end{equation}
and
\begin{equation}
V(N,t_1,t_2) \geq V'(N,t_1,t_2), \forall ]t_1,t_2] \subset ]0,+\infty[,
\end{equation}
then it holds that
\begin{equation}
x_i^{]t_1,t_2]}\geq x_i^{,]t_1,t_2]}, \forall i\in N \text{ and } \forall ]t_1,t_2] \subset ]0,+\infty[.
\end{equation}

\item Stability (STA). A time window value $\Phi$ is symmetric is for all game $(N,V)$, and for all time window $]t_1,t_2]$, 
\begin{equation}
x^{]t_1,t_2]} \in C(N,V(\cdot,t_1,t_2)).
\end{equation}

\item Time separability (TS). A time window value $\Phi$ satisfies the time separability property, if for all game $(N,V)$, $\forall ]t_1,t_3] \subset ]0+\infty[$ it holds that
\begin{equation}
x^{]t_1,t_3]} = x^{]t_1,t_2]} + x^{]t_2,t_3]}, \forall t_2 \in ]t_1,t_3].
\end{equation}

\item Session separability (SS). A time window value $\Phi$ satisfies the session separability property, if for all game $(N,V)$, it holds
\begin{equation}\label{defvs}
\Phi(N,V)=\sum_{s \in S(0,+\infty)} \Phi(N,V^{s}),
\end{equation}
where $V^{s}(F,t_1,t_2)=\chi\left(\{s\},S(t_1,t_2)\right)R(s|F)$.

\end{itemize}

All these properties are relevant in the context of the problem of allocating the revenues generated in a video platform on Internet to stakeholders. The first four properties are related to fairness. First, it seems sensible that all the revenue that is generated by the system is distributed among all those who contributed in some way to generate it. This principle is reflected in the properties of \textit{efficiency} and \textit{null player}. Second, if two stakeholders contribute the same to the generation of revenue in the system, for a matter of justice and non-discriminatory treatment in the distribution, it seems appropriate that both receive the same. This idea of justice is reflected in the property of \textit{symmetry}. Third, if the cooperation of all stakeholders generates more revenue, it seems reasonable that no stakeholder receives less than before when the new larger total revenue is distributed. This criterion of not getting worse if revenue grows is captured by the property of \textit{monotonicity}. Fourth, the allocation of revenue should be done in such a way that no stakeholder or group of stakeholders can at any time complain about the amount of revenue that has been allocated to them, otherwise they could leave the platform. This could lead either to a reduction or to frequent changes in the content offered on the platform, which could cause a loss of users. This principle of avoiding stakeholder complaints and, therefore, of guaranteeing in some way the continuity of the content on the platform is reflected by the property of \textit{stability}. All these properties can be considered basic for any revenue allocation problem, in particular, for those in the context of Internet video platforms such as those described in this paper. However, the other two properties are closely related to the structure of the systems studied in this work. First, it seems reasonable to think in this context that the final allocation of revenue is independent of the time windows applied for its distribution and, therefore, that the final allocation may be the sum of partial allocations of revenue and, at the same time, that future revenue allocations does not depend on the history up to a certain point. This principle also enables us to break down the problem into independent time windows. All these ideas of independence of the final allocation of the specific moments in which the distributions are made are reflected in the property of \textit{time separability}. Second, in the context of Internet video platforms such as those described in this paper, it seems sensible to think that the revenue generated in a session is considered independent of that generated in any other session and, therefore, that the beneficiaries of a session are only those stakeholders who have contributed to generate the revenue of that session. This enables us to break down the allocation problem into many independent allocation problems associated each of them with a different session. These ideas are reflected in the property of \textit{session separability}. In addition, these last two properties allow for easier technical maneuverability in determining the allocation of revenue among stakeholders.

\section{Revenue allocation rules}
There are many ways to define an allocation rule and we can find different rules in the literature. Thus, the use of an allocation rule or another may be based, for example, on the idea of fairness that the manager in charge of making the revenue distribution has.

As mentioned, we can find many allocation rules in the literature, but one of the most prominent rules when we have defined an associated game with the allocation problem is the Shapley value. Furthermore, in this context of Internet video platforms, the Shapley value makes sense due to the particular dynamic usage structure of these platforms. The navigation paths of users in an Internet video platform consist of a succession of events arranged in time. Where each of the events can generate a certain revenue. It is worth asking what part of the revenue generated in each event is attributable to the events that happened previously, evaluating whether that revenue would have occurred if one or more of the previous events had not occurred simply because the videos associated with the events or their channels they were not on the platform. Both the sequential structure of events (associated with players) and the what if question are closely related to the ideas behind the Shapley value \cite{Shapley1953,Algaba2019d}. Therefore, it seems appropriate to use the Shapley value as a distribution rule of the revenue generated in an Internet video platform.

In the next subsections we study how the Shapley value of our games is, and introduce new values which are defined by modifying a bit some relevant characteristic in the structure of the Shapley value for the problems at hand.

\subsection{The Shapley value}
In this subsection we analyze how the Shapley value of the games introduced in this paper is. Thus, we will study the Shapley value for each of the two discrete event dynamic games introduced in Subsection \ref{discreteeventgames}.

We start with the game $V$ which is defined under Assumption \ref{a1} and Assumption \ref{a2-all_or_nothing}. First, note that 
\begin{equation}
V=\sum_{s \in S(0,+\infty)}V^{s}
\end{equation}
Indeed, take $F \subset N$ and $]t_1,t_2] \subset ]0,+\infty[$, then
\begin{align}
V(F,t_1,t_2) & =\sum_{s \in S(t_1,t_2)}V^{s}(F,t_1,t_2) \\
& =\sum_{s \in S(t_1,t_2)}\chi\left(\{s\},S(t_1,t_2)\right)R(s|F) \\
& = \sum_{s\in S(t_1,t_2)} R(s|F), \label{eq37}
\end{align}
where \eqref{eq37} follows from the definitions of $S(t_1,t_2)$, and $\chi\left(\{s\},S(t_1,t_2)\right)$.

Now, since the Shapley value is additive \cite{Shapley1953}, it is session separable,
\begin{equation}\label{sh-session-sep}
Sh(N,V)=\sum_{s \in S(0,+\infty)} Sh(N,V^{s}).
\end{equation}

In view of \eqref{sh-session-sep}, we can focus our analysis on the games $(N,V^{s})$.
\medskip
\begin{theorem}
Given a dynamic game $(N,V) \in DD_{12}^{N}$, and a session $s \in S(0,+\infty)$, the time window Shapley value of the game $(N,V^{s})$ is given by
\begin{align}
&Sh_{i}^{]t_1,t_2]}(N,V^{s})= \nonumber \\
&=\begin{cases}
0, & \text{if } s \notin S(t_1,t_2) \\
\sum_{k=1}^{n_s}\frac{\chi\left(\{i\},P(E^k(s))\right)}{n^p_k(s)} r(e_k(s)),& \text{if } s \in S(t_1,t_2)
\end{cases}.
\end{align}
\end{theorem}
\medskip
\begin{corollary}
Given a dynamic game $(N,V) \in DD_{12}^{N}$, the time window Shapley value of the game $(N,V)$ is given by
\begin{equation}\label{shapleynotilde}
Sh_{i}^{]t_1,t_2]}(N,V)=
\sum_{s \in S(t_1,t_2)} \sum_{k=1}^{n_s}\frac{\chi\left(\{i\},P(E^k(s))\right)}{n^p_k(s)} r(e_k(s)).
\end{equation}
\end{corollary}

First, note that the video website $W$ is represented by three players, $\{w_p,w_s,w_r\}$, so the revenue obtained by $W$ will be the sum of its three representatives. In this sense, $W$ knows how much it obtains from each part of the system. Second, we can observe that the influence of a player in the revenue generation in a particular event is not attenuated over time, i.e., it is the same that an event is at the beginning of the session or just before another event when calculating the  part of revenue generated by the last one that is allocated to the first one.

We continue with the game $\widetilde{V}$ which is defined under Assumption \ref{a1} and Assumption \ref{a3-nothinghappens}. Everything exposed for game $V$ is also valid for game $\widetilde{V}$, simply changing $V^{s}$ for $\widetilde{V}^{s}$, and $R(s|F)$ for $\widetilde{R}(s|F)$. Therefore, we can again focus our analysis on the games $(N,\widetilde{V}^{s})$.
\medskip
\begin{theorem}
Given a dynamic game $(N,V) \in DD_{13}^{N}$, and a session $s \in S(0,+\infty)$, the time window Shapley value of the game $(N,\widetilde{V}^{s})$ is given by
\begin{align}
&Sh_{i}^{]t_1,t_2]}(N,\widetilde{V}^{s})= \nonumber \\
&=\begin{cases}
0, & \text{if } s \notin S(t_1,t_2) \\
\sum_{k=1}^{n_s}\frac{\chi\left(\{i\},P(\widetilde{E}^k(s))\right)}{2} r(e_k(s)),& \text{if } s \in S(t_1,t_2)
\end{cases}.
\end{align}
\end{theorem}
\medskip
\begin{corollary}
Given a dynamic game $(N,\widetilde{V}) \in DD_{13}^{N}$, the time window Shapley value of the game $(N,\widetilde{V})$ is given by
\begin{equation}\label{shapleytilde}
Sh_{i}^{]t_1,t_2]}(N,\widetilde{V})=
\sum_{s \in S(t_1,t_2)} \sum_{k=1}^{n_s}\frac{\chi\left(\{i\},P(\widetilde{E}^k(s))\right)}{2} r(e_k(s)).
\end{equation}
\end{corollary}

First, note that the video website $W$ is again represented by three players, $\{w_p,w_s,w_r\}$, hence we have the same as the previous game. Second, in this case we only take into account the platform, $w_p$, and the corresponding event when allocating the revenue, i.e., no other information related to the moment an event occurred is consider in the definition of the $\widetilde{V}$ games.

Some final comments about the Shapley value for these games are the following. First, the timewise Shapley values for the games $V$ and $\widetilde{V}$ can easily be derived by considering the time window $]0,\infty[$. The games $V$ and $\widetilde{V}$ are timewise convex, therefore their timewise Shapley values belong to their timewise cores. This means that the timewise Shapley value satisfies stability. Likewise,  the time window Shapley value satisfies all properties listed in Section \ref{properties}. Finally, for the game $\widetilde{V}$, many more things cannot be said, but as we will see in the following subsections for the game $V$, we can take into account other elements when defining allocation rules. These will be variations on the Shapley value structure of this subsection, i.e., they are defined based on the Shapley value structure but considering more information.

\subsection{Another non-attenuated over time rule}
First, we start with allocation rules in which the revenue allocation to the players do not depend on the moment they appear in a session, i.e., they are non-attenuated over time. We observe that this approach can only be taken in the $V$ games because very little information is consider in the definition of the $\widetilde{V}$ games in order to define new allocation rules based on the structure of the Shapley vale.

Note that the information not used in the definition of the Shapley value is the number of times that events belonging to a player are viewed. If we take into account that information, we can define a new allocation rule based on the structure of the Shapley value in the following way. First, for games $(N,V^{s})$, the new time window rule is given for every $i \in W \cup C$ by

\begin{align}\label{discreteshapley}
&ES_{i}^{]t_1,t_2]}(N,V^{s})= \nonumber \\
&=\begin{cases}
0, & \text{if } s \notin S(]t_1,t_2]) \\
\sum_{k=1}^{n_s}\frac{|E_i^{k}(s)|}{k+1} r(e_k(s)),& \text{if } s \in S(]t_1,t_2])
\end{cases}
\end{align}
where $E_i^{k}(s)\subset E^{k}(s)$ is the set of all events in $E^{k}(s)$ that belong to $i$.

Now, adding \eqref{discreteshapley} for every session in the time window $]t_1,t_2]$, we define the time window event-Shapley rule as

\begin{equation}
ES_{i}^{]t_1,t_2]}(N,V)=\sum_{s \in S(t_1,t_2)} \sum_{k=1}^{n_s}\frac{|E_i^{k}(s)|}{k+1} r(e_k(s)),
\end{equation}
where $E_i^{k}(s)\subset E^{k}(s)$ is the set of all events in $E^{k}(s)$ that belong to $i$.

Note that the main difference of this new rule with regards to the time window Shapley value is that the former takes into account the number of times that a player appears in each sessión. The time window event- Shapley rule inherits all properties of the time window Shapley but symmetry. Likewise, we can define the timewise event-Shapley rule by simply considering the time window $]0,+\infty[$.

With respect to the properties, the time window Shapley value satisfies all the properties listed in Section \ref{properties}, while the rule introduced in this subsection does not satisfy the symmetry property, so the definition of this rule does not seem to improve the time window Shapley value. However, there is a property that does not satisfy the time window Shapley value, but the rule introduced in this subsection does satisfy: the property of non-manipulability in the context of discrete event dynamic games. This property says that if a channel makes the decision to divide into several channels or several channels make the decision to join into one, the result for them would be the same. Formally this property is as follows:
 
First, given a game $(N,V) \in DD_{12}$, we denote by $v(i)$ the set of all videos of channel $i$.
 
\begin{itemize}
\item Non-manipulability (NM). A time window value $\Phi$ satisfies the non-manipulability property, if given two games $(N,V), (N',V') \in DD_{12}$, such that $N' \subset N$ and there is a channel $i \in N'$ such that $v'(i)=v(i) \cup v(N\backslash N')$, and for all $j\in N'\backslash {i}$, $v'(j)=v(j)$, then
\begin{equation}
\phi_i(N',V') =\phi_i(N,V)+\sum_{j\in N\backslash N'} \phi_j(N,V).
\end{equation}
\end{itemize}

Note that in the definition of the property of non-manipulability we assume that the sessions and the revenues generated by each event are the same in both games. The only difference between the games $(N,V)$ and $(N',V')$ is the owners of the events (videos). Moreover, this property is only given for the context of discrete event dynamic games, i.e., it is not a general property for dynamic games given in Subsection \ref{preliminar}. 

A final comment is that in the context of games in $DD_{13}$, the time window Shapley value does also satisfy the property of non-manipulability.

\subsection{A family of attenuated over time rules}
In the previous subsections, we have introduced rules in which when distributing the revenue generated by an event among all the players, it was only taken into account if players were before or after the event to consider whether they participated in the distribution of that revenue. However, the distance of the players to the event at hand was not relevant when determining the allocation of the revenue. Nevertheless, it seems reasonable to think that if a player is very far from the event of which the revenue is being distributed, her influence will be increasingly weaker, i.e., the farther the less influence. Therefore, when distributing the revenue generated in a particular event, it seems reasonable that this should be taken into account. If we accept this, then we have to decide how a player's influence weakens as he moves away from the event from which the revenue is being shared. For this purpose, we first introduce the attenuation functions.
\medskip
\begin{definition}
A discrete attenuation function $\alpha$ is a map from the set of the non negative integers, $\mathbb{Z}_+$, to the interval $[0,1]$, such that
\begin{enumerate}
\item $\alpha (0) =1$, and
\item $\alpha (k) \leq \alpha (k-1), \forall k \geq 1.$
\end{enumerate}
\end{definition}
\medskip

The alpha function describes how the influence of the players is attenuated with the distance to the event at hand. First, for games
$(N,V^{s})$, the new time window rule with attenuation function $\alpha$ is given for every $i \in W \cup C$ by

\begin{align}\label{alphashapley}
&ES_{i}^{\alpha,]t_1,t_2]}(N,V^{s})= \nonumber \\
&\begin{cases}
0, & \text{if } s \notin S(]t_1,t_2]) \\
\sum_{k=1}^{n_s}\frac{\sum_{e_l(s)\in E_{i}^{k}(s)}\delta_k^{\alpha}(k-l)\alpha(k-l)}{\sum_{j=0}^{k}\delta_k^{\alpha}(k-j)\alpha(k-j)} r(e_k(s)),& \text{if } s \in S(]t_1,t_2])
\end{cases}
\end{align}
where $\delta_k^{\alpha}(k)= \frac{1}{\alpha(k)}$, and 1 otherwise; and $E_i^{k}(s)\subset E^{k}(s)$ is the set of all events in $E^{k}(s)$ that belong to $i$.

If we look carefully at \eqref{alphashapley}, we observe that the summand corresponding to player $w_p$ is always 1 both in the denominator and when it appears in the numerator. With this we assume that the distance between any event and the platform is always zero, which seems reasonable since the event can occur because the platform is working at that moment.

Now, adding \eqref{alphashapley} for every session in the time window
$]t_1,t_2]$, we define the time window $\alpha$-Shapley rule as

\begin{align}
& ES_{i}^{\alpha,]t_1,t_2]}(N,V^{s})= \nonumber \\
&\sum_{s \in S(t_1,t_2)}\sum_{k=1}^{n_s}\frac{\sum_{e_l(s)\in E_{i}^{k}(s)}\delta_k^{\alpha}(k-l)\alpha(k-l)}{\sum_{j=0}^{k}\delta_k^{\alpha}(k-j)\alpha(k-j)} r(e_k(s)),
\end{align}
where $\delta_k^{\alpha}(k)= \frac{1}{\alpha(k)}$, and 1 otherwise; and $E_i^{k}(s)\subset E^{k}(s)$ is the set of all events in $E^{k}(s)$ that belong to $i$.

Obviously, we can define many different attenuation functions, but we now focus on a special type: the family of exponential attenuation functions. One member of this family is defined as follows:
\begin{equation}
\beta(k)=\theta^{k}, \forall k \in \mathbb{Z}_+,
\end{equation}
where $\theta \in [0,1]$. 

We assume that $0^0=1$. Thus, when $\theta = 0$, we obtain the Shapley value given by \eqref{shapleytilde}; and when $\theta = 1$, we obtain the Shapley value given by \eqref{discreteshapley}. Therefore, we have two poles, from all players are irrelevant except the platform and the player who owns the corresponding event, until all players are equally relevant to the event at hand. This is interesting because it allows us to have both extremes in the same family. Nevertheless, the Shapley value given by \eqref{shapleynotilde} cannot be obtained with any $\theta \in [0,1]$. In fact, it cannot be obtained for any attenuation function $\alpha$. Therefore, on the one hand we have the Shapley value given by \eqref{shapleynotilde}, and on the other hand we have the family of exponential attenuation rules.

Finally, all attenuated over time rules introduced in this subsection satisfy the same properties as the time window event-Shapley rule. Therefore, they are also good candidates to be considered when allocating the revenues generated in a video website.


\section{Algorithm for computing the allocation rules}
It is well known that one of the biggest handicaps of solutions in
cooperative games is their computational complexity, generally of
the exponential type, i.e., $O(a^{N}),\:a>1$, where $N$ is the number
of players involved in the cooperative game. However, as we have seen
previously, the structure of the problem allows us to obtain allocations
of the revenues in an intuitive and simple way, in particular, simple expressions have been obtained for the Shapley value and Shapley-like allocation rules, which allow their calculations in an efficient way. A more detailed analysis of the formulas obtained for the Shapley value and the Shapley-like allocation rules allows us to conclude that, given $S(t_1,t_2)$, their computational complexities are of polynomial type, in particular at most $O(SK^{2})$, where $S$ is the number of sessions in $S(t_1,t_2)$ and $K$ is the maximal number of events of a session in $S(t_1,t_2)$. Therefore, a great improvement is obtained from the computational point of view, making these solutions suitable, not only from the perspective of the fairness and their good behavior with respect to a large number of properties, but also from the point of view of their computation in a reasonable time.

Below we show how a possible algorithm would work, and its computational complexity, in the case of applying an attenuated over time allocation rule. For the rest of the values and rules introduced in this paper, the algorithm would be basically the same, changing only the first step, which corresponds to the formation of the relevance matrix.

When a session $s$ starts, a matrix is generated that gathers the relevance of an event on the revenue generated in successive events. For example, for a discrete attenuation function $\alpha$ and the $k$-th event of the session, the following column array would be added to the relevance matrix:

\begin{equation}
\begin{array}{c}
1 \\
\alpha\left({k-1}\right)\\
\alpha\left({k-2}\right)\\
\vdots \\
\alpha\left(0\right)
\end{array}
\end{equation}
 
At the same time, we generate a revenue column array for the events, where the $k$-th cell corresponds to the revenue generated during the $k$-th event of the session.
 
When the session ends, the non-assigned cells of the relevance matrix are completed with 0's. Thus, at the end of the session we have the following relevance matrix and revenue vector:

\begin{equation}
\left(\begin{array}{ccccc}
1 & 1 & \cdots & 1 & 1 \\
\alpha(0) &\alpha(1) & \cdots & \alpha(n_s-2) & \alpha(n_s-1) \\
0 & \alpha(0) &  \cdots & \alpha(n_s-3) & \alpha(n_s-2) \\
0 & 0 & \cdots & \alpha(n_s-4) & \alpha(n_s-3) \\
\vdots & \vdots & \vdots & \vdots & \vdots \\
0 & 0 & \cdots & 0 & \alpha(0) 
\end{array}\right),
\end{equation}
\begin{equation}
\left(\begin{array}{c}
r(e_1(s))\\
\vdots\\
r(e_{n_s}(s)
\end{array}\right).
\end{equation}

Therefore, the relevance matrix has $n_s+1$ rows and $n_s$ columns, and the revenue vector has $n_s$ rows and only one column.

This procedure takes a number of elementary operations proportional to $n_s(n_s+1)$, where $n_s$ is the number of events of the session.

The second step consists in dividing each of the cells of the relevance matrix by the sum of the values in the cells of the column to which it belongs. This gives us a matrix of weights. For example, the cell $(i,j)$ of the weight matrix would be given by
\begin{equation}
\frac{r_{ij}}{\sum_hr_{hj}},
\end{equation}
where $r_{hj}$ is the value for the cell $(h,j)$.

This procedure again takes a number of elementary operations proportional to $n_s(n_s+1)$.

The third step simply consists in multiplying the weight matrix by the revenue vector. This also takes a number of elementary operations proportional to $n_s(n_s+1)$. The final step is to assign the cells of the last matrix to each player, where possibly more than one cell may correspond to the same player. This takes a number of elementary operations proportional to $n_s+1$.

Now it is easy to conclude that this algorithm has a computational complexity $O(n_s^2)$. Therefore, when we consider a time window $]t_1,t_2]$, the computational complexity will be at most $O(SK^{2})$, where $S$ is the number of sessions in $S(t_1,t_2)$ and $K$ is the maximal number of events of a session in $S(t_1,t_2)$, as noted above.

This algorithm shows that the values and allocation rules introduced in this paper can be computed in real-time and, therefore, can be adapted to the time dynamics of use of the video website. Therefore, our revenue allocation approach based on cooperative games is suitable for the real dynamic situation describing the operation of a video website.


\section{Illustrative Examples}
Although from the structure of the values and allocation rules introduced in this paper it can be deduced who benefits from what rules. In this section we provide some examples to illustrate the differences among the different values and rules introduced. 

Let us consider a time window $]t_1,t_2]$ during which the three sessions given in Table 1 end, and to simplify we assume that the events associated with player 1 generate a revenue of 3 m.u., the events associated with player 2 of 6 m.u., the events associated with player 3 of 9 m.u., the events associated with player $w_r$ of 1 m.u., and the rest of events nothing at all.
\begin{table}
\begin{center}
\begin{tabular}{|c|ccccccccccc|}\hline
          & $e_0$ & $e_1$ & $e_2$ & $e_3$ & $e_4$ & $e_5$ & $e_6$ & $e_7$ & $e_8$ & $e_9$ & $e_{10}$ \\ \hline    
 $s_1$ & $w_p$ & $w_s$ & 1 & $w_r$ & 2 & $w_r$ & 1 & $w_r$ & 3 & $w_r$ & 2 \\
 $s_2$ & $w_p$ & 3 & $w_r$ & 1 & 1 & $w_r$ & 2 & $w_r$ & 2 & $w_s$ & -- \\
 $s_3$ & $w_p$ & $w_s$ & 2 & 2 & $w_r$ & 3 & $w_r$ & 1 & 1 & -- & -- \\ \hline 
\end{tabular}
\end{center}
\caption{Events and players associated with each session.}
\end{table}

We start with the time window Shapley value of the game $V$. In this case, for example for session $s_3$, the revenue allocation is given by
\medskip

\begin{center}
{\scriptsize
\begin{tabular}{c|ccccccccc|c}
$r(e_k(s_1))$ & 0 & 0 & 6 & 6 & 1 & 9 & 1 & 3 & 3 & Total \\ \hline
$w_p$       & 0 & 0 & 2 & 2 & 0.25 & 1.8 & 0.2 & 0.5 & 0.5 & 7.25\\
$w_s$       & 0 & 0 & 2 & 2 & 0.25 & 1.8 & 0.2 & 0.5 & 0.5 & 7.25 \\
$w_r$       & 0 & 0 & 0 & 0 & 0.25 & 1.8 & 0.2 & 0.5 & 0.5 & 3.25\\
1               & 0 & 0 & 0 & 0 & 0  & 0 & 0 & 0.5 & 0.5 & 1 \\
2               & 0 & 0 & 2 & 2 & 0.25 & 1.8 & 0.2 & 0.5 & 0.5 & 7.25 \\
3               & 0 & 0 & 0 & 0 & 0  & 1.8 & 0.2 & 0.5 & 0.5 & 3 \\ \hline
\end{tabular}}
\end{center}

\medskip
The time window Shapley value and the revenue generated by each player in the time window $]t_1,t_2]$ are

\medskip

\begin{center}
{\footnotesize
\begin{tabular}{c|cccccc}
Players & $w_p$ & $w_s$ & $w_r$ & 1 & 2 & 3 \\ \hline
$Sh_i^{]t_1,t_2]}(N,V)$ & 22.55 & 13.37 & 13.05 & 11.47 & 14.72 & 14.85 \\
Revenues & 0 & 0 & 9 & 18 & 36  & 27 \\ \hline
\end{tabular}}
\end{center}

\medskip

We now consider the following members of the exponential attenuation family of rules: $\theta = 0, \frac{1}{4}, \frac{1}{2},\frac{3}{4},1$, which correspond to the time window Shapley value of the game $\widetilde{V}$, the time window $\frac{a}{b}$-Shapley rules, and the time window event-Shapley value, respectively. This choice allows us to show conveniently all types of rules introduced in this paper. Below we can see the revenue allocation applying the mentioned rules and the direct revenue generated during the time window at hand.

\medskip

\begin{center}
{\footnotesize
\begin{tabular}{l|rrrrrr}
Players & $w_p$ & $w_s$ & $w_r$ & 1 & 2 & 3 \\ \hline
$Sh_i^{]t_1,t_2]}(N,V)$ & 22.55 & 13.37 & 13.05 & 11.47 & 14.72 & 14.85 \\ \hline
$\theta = 0$ & 45.00 & 0.00 & 4.50	 & 9.00 & 18.00 & 13.50 \\
$\theta = 1/4$ & 39.42	& 1.26	& 9.86 & 9.42 & 17.13 & 12.91 \\
$\theta = 1/2$ & 32.75 & 3.05 & 14.33 & 10.60 & 16.76 & 12.49 \\
$\theta = 3/4$ & 25.38 & 6.06 & 17.43 & 12.27 & 16.74	& 12.12 \\
$\theta = 1$ & 18.87 & 10.87 &	18.11 & 13.64 & 16.53 &	12.00 \\ \hline
Revenues & 0.00 & 0.00 & 9.00 & 18.00 & 36.00  & 27.00 \\ \hline
\end{tabular}}
\end{center}

\medskip

We can observe that the revenue allocated to player 1 increases and the revenues allocated to players 2 and 3 decrease from $\theta=0$ to $\theta=1$. This happens because player 1 is more times located before the other two and this favors her when the influence is not attenuated with the distance in time. The same goes for $w_s$ and $w_r$. The revenue allocated to player $w_p$ decreases from $\theta=0$ to $\theta=1$ because her distance to the event is always zero, so there are others that increase their influence hurts her. We also observe that the time window Shapley value of the game $V$ has a quite different behavior than the other allocation rules.

Next, we analyze by simulation the behavior of the allocation rules for sessions with 5, 10, 15 and 20 events. For this, we consider time windows with 100 sessions, all of them with the same number of events. And we replicate this 10 times. In total we consider 10 time windows with 100 sessions each, for sessions of 5, 10, 15 and 20 events, respectively.

The behavior of the users on the video website is described by the following vector $I$ that indicates where users start,
\begin{equation}\label{initialvector}
\left(
\begin{array}{l}
Pr(w_s) \\
Pr(w_r) \\
Pr(1) \\
Pr(2) \\
Pr(3) 
\end{array}
\right) =
\left(
\begin{array}{l}
0.25 \\
0.13 \\
0.25 \\
0.25 \\
0.12
\end{array}
\right)
\end{equation}
and the following transition matrix $M$ that shows how users change from one event to another, remember that the events are identified with the players:

\begin{equation}\label{transitionmatrix}
\begin{tabular}{l|ccccc|}
       & $w_s$ & $w_r$ & 1 & 2 & 3 \\ \hline
$w_s$ & 0.10 & 0.40 & 0.20 & 0.20 & 0.10 \\
$w_r$ & 0.00 & 0.00 & 0.40 & 0.40 & 0.20 \\
1     & 0.10 & 0.50 & 0.40 & 0.00 & 0.00 \\
2     & 0.10 & 0.50 & 0.00 & 0.40 & 0.00 \\
3     & 0.10 & 0.70 & 0.00 & 0.00 & 0.20
\end{tabular}
\end{equation}

Vector $I$ given by \eqref{initialvector} and transition matrix $M$ given by \eqref{transitionmatrix} show that Channels 1 and 2 have approximately the double of videos than Channel 3, and the interest of videos for users is the same, i.e., there are no videos from one channel more attractive than videos from another channel.

Finally the direct revenues generated by each event depending on to whom it belongs are the same as the previous example, i.e., the events associated with player 1 generate a revenue of 3 m.u., the events associated with player 2 of 6 m.u., the events associated with player 3 of 9 m.u., the events associated with player wr of 1 m.u., and the rest of events nothing at all.

Now, we run the simulation for the time window Shapley value of the game $V$, and as before for the following members of the exponential attenuation family of rules: $\theta = 0, \frac{1}{4}, \frac{1}{2}, \frac{3}{4}, 1$, which correspond to the time window Shapley value of the game $\widetilde{V}$, the time window $\frac{a}{b}$-Shapley rules, and the time window event-Shapley value, respectively. The results of the simulation are presented in terms of the proportion of the total revenue each player obtains.

\begin{figure}
\begin{center}
\includegraphics[width=5.5in,height=4.5in,clip,keepaspectratio]{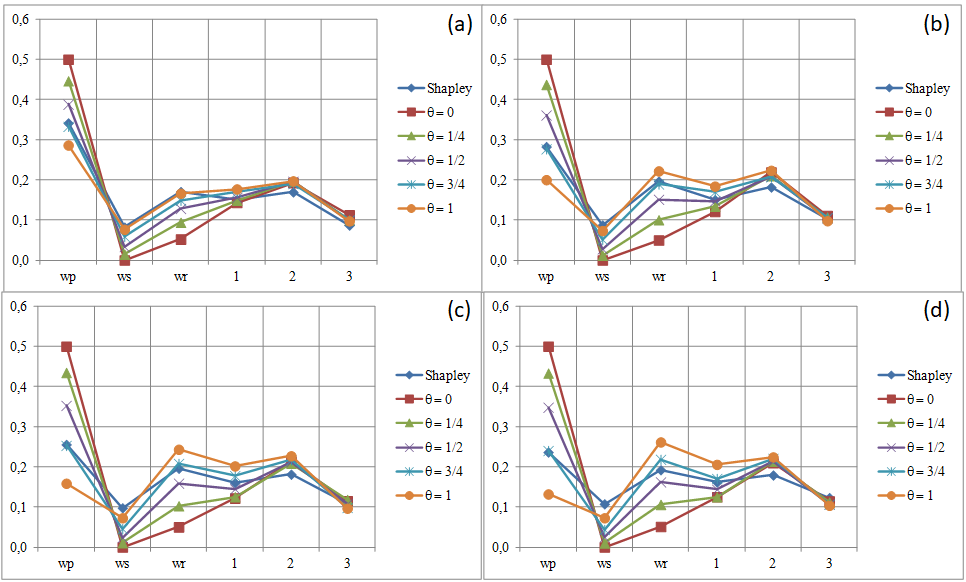}
\end{center}
\caption{Average allocations for sessions with (a) 5, (b) 10, (c) 15, and (d) 20 events.}
\label{revenuedistribution}
\end{figure}

In Fig. \ref{revenuedistribution}, we show the proportion of the revenues that each player gets for sessions of size from 5 to 20. We observe that for channels is more profitable to have more videos than fewer videos but more revenue generators. For example, if we compare channels 1 and 3, we know that Channel 1 has the double of videos than Channel 3, but the videos of Channel 3 generates the triple of revenues than the ones of Channel 1, however Channel 1 always gets on average a higher allocation than Channel 1 in all scenarios. Likewise, we observe that the shape of the average allocation distribution is similar in all scenarios.

\begin{figure}
\begin{center}
\includegraphics[width=5.0in,height=4.0in,clip,keepaspectratio]{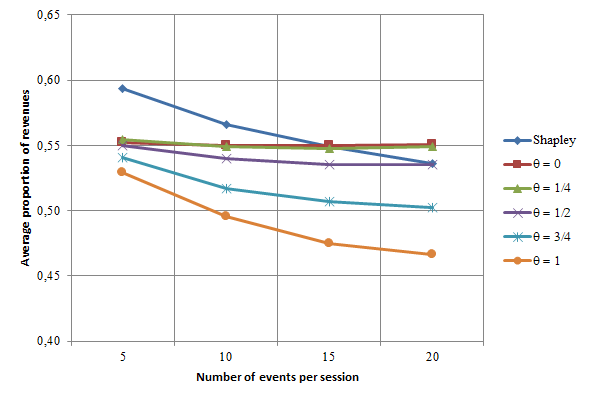}
\end{center}
\caption{Average proportion of revenues allocated to the platform.}
\label{platformallocation}
\end{figure}

On the other hand, the average proportion of revenues allocated to the platform ($w_p, w_r, w_s$) decreases with the number of events of sessions when the rules introduced in this paper are implemented (see Fig. \ref{platformallocation}). Moreover, we can observe that the time window $0$-Shapley rule and the time window $\frac{1}{4}$-Shapley rule seem the more beneficial for the platform, in general. Likewise, the most sensitive rules to the length of the sessions are the time window Shapley value and the Shapley rules with $\theta > \frac{1}{2}$ (see Fig. 4).

\section{Conclusion}
In this paper we have analyzed a video website system as a multi-party market. In particular, we have studied how to distribute the revenue generated in the system among all the stakeholders that contribute to it, specifically the platform and content creators. For this purpose, we have introduced a dynamic model of revenue generation and associated with it, we have defined two different dynamic games according to several assumptions. For these games we have determined their Shapley values. Based on the structure of the Shapley value, we have introduced new allocation rules. In particular, another non-attenuated over time rule and a family of attenuated over time rules. All rules introduced in this paper, including the Shapley value, satisfy a set of fairness properties which makes them reasonably suitable to be applied in the revenue allocation problem. 

On the other hand, another novel aspect of this work, apart from the already mentioned dynamic model of revenue generation in an Internet video system, the associated dynamic games and the introduced allocation rules, is the fact of combining the ideas of the dynamic cooperative games and attribution problems to address the problem of revenue distribution in an Internet video system, considering the influence of each of the events on the navigation path of users in the system and not just the event concrete that generates revenue.

Moreover, a polynomial time algorithm has been implemented to compute all the allocation rules introduced in the paper, which shows that the results are fully applicable to a real system. In addition, some simulation computational experience is presented to illustrate the performance of the introduced allocation rules.

Finally, we would like to draw attention to the rules that incorporate attenuation over time, because in this work we have considered attenuation taking into account how many events have elapsed up to a given instant, but not how long. At this point, it seems interesting to consider other models that take into account how long each of the events lasts because in many cases it could not be the same as an event lasting 5 minutes or 15 minutes. Therefore, it would be interesting for further research to study dynamic models that take into account the duration of each event or even continuous dynamic models.


\section{Appendix: proofs}
\subsection{Proof of Proposition 1}
First note that if the number of addends of a sum is empty, then the sum is $0$. From now on, for the sake of simplicity, we denote $max\{h| P(E^h(s)) \subset F\}$, by $h^{max}(F)$.

Let $(N,V) \in DD_{12}$, and $F \subset G \subset N\backslash \{i\}$. By \eqref{DD12a} and \eqref{DD12b}, it holds that
\begin{align}
&V(F \cup \{i\},t) = \sum_{s\in S(t)} \sum_{k\leq h^{max}(F \cup \{i\})}r(e_k(s)), \\
&V(F,t) = \sum_{s\in S(t)} \sum_{k\leq h^{max}(F)}r(e_k(s)).
\end{align}

We know that if $\{e_0(s),\ldots,e_h(s)\} \subset F$, then $\{e_0(s),...,e_h(s)\} \subset F \cup \{i\}$. Therefore, for each session $s$, the difference between the two maximal chains of events contained in $F \cup \{i\}$ and $F$, respectively, is $\{e_{k^0_i}(s),\ldots,e_h(s)\}$, where $e_{k^0_i}(s)$ is the first event in the maximal chain contained in $F \cup \{i\}$ such that $e_{k^0_i}(s) = i$. Therefore,
\begin{align}
&V(F \cup \{i\},t)-V(F,t) = \nonumber \\
& \sum_{s\in S(t)} \sum_{k^0_i\leq k\leq h^{max}(F \cup \{i\})}r(e_k(s)).
\end{align}

In the same way, it holds that

\begin{align}
&V(G \cup \{i\},t)-V(G,t) = \nonumber \\
& \sum_{s\in S(t)} \sum_{k^0_i\leq k\leq h^{max}(G \cup \{i\})}r(e_k(s)).
\end{align}

Now, since $F\cup \{i\} \subset G\cup \{i\}$, for each session $s$, the maximal chain of events associated with $F\cup \{i\}$ is contained in the corresponding associated with $G\cup \{i\}$, therefore, it is clear that
\begin{equation}
V(F \cup \{i\},t)-V(F,t) \leq V(G \cup \{i\},t)-V(G,t).
\end{equation}

Now, let $(N,V) \in DD_{13}$, and $F \subset G \subset N\backslash \{i\}$. By \eqref{DD13a} and \eqref{DD13b}, it holds that
\begin{align}
&\widetilde{V}(F \cup \{i\},t) = \sum_{s\in S(t)} \sum_{k : \{e_0(s), e_k(s)\} \subset F \cup \{i\}}r(e_k(s)), \\
&\widetilde{V}(F,t) = \sum_{s\in S(t)} \sum_{k : \{e_0(s), e_k(s)\} \subset F}r(e_k(s)).
\end{align}

Therefore, if $w_p \in F$, then
\begin{equation}
\widetilde{V}(F \cup \{i\},t) -\widetilde{V}(F,t) = \sum_{s\in S(t)}\sum_{k : \{e_0(s), e_k(s)\} \subset \{w_p, i\}}r(e_k(s)),
\end{equation}
otherwise, $\widetilde{V}(F \cup \{i\},t) -\widetilde{V}(F,t) =0$.

Similarly, it holds that if $w_p \in G$, then
\begin{equation}
\widetilde{V}(G \cup \{i\},t) -\widetilde{V}(G,t) = \sum_{s\in S(t)}\sum_{k : \{e_0(s), e_k(s)\} \subset \{w_p, i\}}r(e_k(s)),
\end{equation}
otherwise, $\widetilde{V}(G \cup \{i\},t) -\widetilde{V}(G,t) =0$.

Therefore,
\begin{equation}
\widetilde{V}(F \cup \{i\},t) -\widetilde{V}(F,t) =\widetilde{V}(G \cup \{i\},t) -\widetilde{V}(G,t).
\end{equation}


\subsection{Proof of Proposition 2}
Let $(N,V) \in DD_{12}$, $F \subset N$, and $t_1 < t_2 \leq t_3$. By \eqref{DD12a} and \eqref{DD12b}, it holds that
\begin{equation}
V(F,t_1,t_3) = \sum_{s\in S(t_1,t_3)} \sum_{k\leq h^{max}(F)}r(e_k(s)).
\end{equation}

By definition of $S$, $S(t_1,t_3) = S(t_1,t_2) \cup S(t_2,t_3)$, and $S(t_1,t_2) \cap S(t_2,t_3)=\varnothing$.  Therefore,
\begin{align}
&V(F,t_1,t_3) = \sum_{s\in S(t_1,t_2) \cup S(t_2,t_3)} \sum_{k\leq h^{max}(F)}r(e_k(s))=\\
&\sum_{s\in S(t_1,t_2)} \sum_{k\leq h^{max}(F)}r(e_k(s)) + 
\sum_{s\in S(t_2,t_3)} \sum_{k\leq h^{max}(F)}r(e_k(s)) = \\
&V(F,t_1,t_2) + V(F,t_2,t_3).
\end{align}

For $DD_{13}$, the proof is, {\it mutatis mutandis}, completely analogous to the proof for $DD_{12}$.


\subsection{Proof of Theorem 1}
Let $(N,V) \in DD_{12}^{N}$, a session $s \in S(0,+\infty)$, and the time window $]t_1,t_2]$. By Definition \ref{twsh}, the time window Shapley value of $(N,V^s)$ is given by
\begin{align}
&Sh_i^{]t_1,t_2]}(N,V^s)= \nonumber \\
&\sum_{F \subset N\backslash \{i\}}{\textstyle\frac{|F|!|N\backslash (F\cup \{i\})|!}{|N|!}}\left(V^s(F\cup \{i\},t_1,t_2)-V^s(F,t_1,t_2)\right).
\end{align}

First note that if $s \notin S(t_1,t_2)$, then by definition of $V^s$ (see \eqref{defvs}) it is easy to prove that $Sh_i^{]t_1,t_2]}(N,V^s) =0, \forall i \in N$. 

We now consider that $s \in S(t_1,t_2)$ and $s$ has exactly $n_s+1$ events, $e_0(s),e_1(s),\ldots e_{n_s}(s)$. In order to prove the result, we consider the following alternative expression of the Shapley value \cite{Shapley1953} adapted to the dynamic games in this paper:
\begin{align}
&Sh_i^{]t_1,t_2]}(N,V^s)=\nonumber \\
&\frac{1}{|N|!}\sum_{\pi \in \Pi}\left(V^s(P(\pi,i) \cup \{i\},t_1,t_2)-V^s(P(\pi,i),t_1,t_2)\right),
\end{align}
where $\Pi$ is the set of all the permutations of players in $N$, and $P(\pi,i)$ is the set of players that precede $i$ in the permutation $\pi$.

Now, we look at the events in session $s$, given by $e_0(s),e_1(s),\ldots e_{n_s}(s)$, and analyze when in a permutation $\pi$ a player $i$ will marginally contribute with the revenues associated with an even $e_k(s)$. The answer is that such player $i$ will contribute if the following two conditions are met:
\begin{enumerate}
\item There is an event before $k$ that belongs to player $i$, i.e., there exists $k' < k$ such that $P(e_{k'}(s))=i$.
\item The permutation $\pi$ is is such that player $i$ appears in it after all the players other than her who have associated all the events before event $k$ in the permutation $\pi$.
\end{enumerate} 

These two conditions can only be satisfied by those players who have some event of session $s$ associated with them before event $k$ occurs. Now the question is in how many permutations the two previous conditions will be fulfilled for a certain player. The answer is in exactly $\frac{1}{n^p_k(s)}$ of all permutations, where $E^k(s) = \{e_0(s),e_1(s),\ldots e_k(s)\}$. Thus, every revenue $r(e_k(s))$ is distributed equally among the players in $P(E^k(s))$. Therefore, when $s \in S(t_1,t_2)$, it holds that
\begin{equation}
Sh_{i}^{]t_1,t_2]}(N,V^{s})= 
\sum_{k=1}^{n_s}\frac{\chi\left(\{i\},P(E^k(s))\right)}{n^p_k(s)} r(e_k(s)), \forall i \in N.
\end{equation}


\subsection{Proof of Corollary 1}
This proof follows directly from Theorem 1, \eqref{DD12a}, \eqref{DD12b}, and the additivity of the Shapley value \cite{Shapley1953}.


\subsection{Proof of Theorem 2}
Let $(N,V) \in DD_{13}^{N}$, a session $s \in S(0,+\infty)$, and the time window $]t_1,t_2]$. By Definition \ref{twsh}, the time window Shapley value of $(N,\widetilde{V}^s)$ is given by
\begin{align}
&Sh_i^{]t_1,t_2]}(N,\widetilde{V}^s)= \nonumber \\
&\sum_{F \subset N\backslash \{i\}}{\textstyle\frac{|F|!|N\backslash (F\cup \{i\})|!}{|N|!}}\left(\widetilde{V}^s(F\cup \{i\},t_1,t_2)-\widetilde{V}^s(F,t_1,t_2)\right).
\end{align}

First note that if $s \notin S(t_1,t_2)$, then by definition of $\widetilde{V}^s$ (see \eqref{defvs}) it is easy to prove that $Sh_i^{]t_1,t_2]}(N,\widetilde{V}^s) =0, \forall i \in N$. 

We now consider that $s \in S(t_1,t_2)$ and $s$ has exactly $n_s+1$ events, $e_0(s),e_1(s),\ldots e_{n_s}(s)$. By definition of $\widetilde{V}^s$, \eqref{DD13a} and \eqref{DD13b}, it holds that
\begin{align}
&Sh_i^{]t_1,t_2]}(N,\widetilde{V}^s)= \nonumber \\
&\sum_{F \subset N\backslash \{i\}}{\textstyle\frac{|F|!|N\backslash (F\cup \{i\})|!}{|N|!}}(\sum_{k : P(\widetilde{E}^k(s)) \subset F \cup \{i\}}r(e_k(s))- \nonumber \\
&\sum_{k : P(\widetilde{E}^k(s)) \subset F}r(e_k(s))) = \nonumber \\[0.5cm]
&\sum_{\substack{F \subset N\backslash \{i\} \\[0.1cm] w_p \in F}}{\textstyle\frac{|F|!|N\backslash (F\cup \{i\})|!}{|N|!}}\sum_{k : P(\widetilde{E}^k(s)) \subset \{w_p, i\}}r(e_k(s)) = \nonumber \\
& \frac{1}{2}\sum_{k : P(e_k(s)) = i}r(e_k(s)) = \sum_{k=1}^{n_s}\frac{\chi\left(\{i\},P(\widetilde{E}^k(s))\right)}{2} r(e_k(s)),
\end{align}
where $\widetilde{E}^k(s)=\{e_0(s),e_k(s)\}$.


\subsection{Proof of Corollary 2}
This proof follows directly from Theorem 2, \eqref{DD13a}, \eqref{DD13b}, and the additivity of the Shapley value \cite{Shapley1953}.


\section*{Acknowledgment}
This work is part of the R\&D\&I project grant PGC2018-097965-B-I00, funded by MCIN/ AEI/10.13039/501100011033/ and by "ERDF A way of making Europe"/EU. The authors are grateful for this financial support. Financial suport from the Generalitat Valenciana under the project PROMETEO/2021/063 is also acknowledged.

\end{document}